\documentclass[12pt]{article}

\usepackage{amsmath}
\usepackage{amssymb}
\usepackage{graphicx}

\def\Re{{\cal R \mskip-4mu \lower.1ex \hbox{\it e}\,}}
\def\Im{{\cal I \mskip-5mu \lower.1ex \hbox{\it m}\,}}
\def\ie{{\it i.e.}}
\def\tev{\,{\rm TeV}}

\def\mev{\,{\rm MeV}}
\def\to{\rightarrow}

\newskip\zatskip \zatskip=0pt plus0pt minus0pt
\def\matth{\mathsurround=0pt}

\def\gsim{\mathrel{\mathpalette\atversim>}}
\def\atversim#1#2{\lower0.7ex\vbox{\baselineskip\zatskip\lineskip\zatskip
  \lineskiplimit 0pt\ialign{$\matth#1\hfil##\hfil$\crcr#2\crcr\sim\crcr}}}

\renewcommand{\thefootnote}{\fnsymbol{footnote}}

\begin{document}

\bibliographystyle{h-physrev}

 \begin{titlepage} 
\rightline{\vbox{\halign{&#\hfil\cr
&SLAC-PUB-10005\cr
&June 2003\cr}}}
%\vspace{1in} 
\vspace{.7in} 
\begin{center}

{\Large\bf
Flavor Constraints on Split Fermion Models}
\footnote{Work supported by the Department of 
Energy, Contract DE-AC03-76SF00515}
%and W-7405-Eng-82.} 
\medskip

\normalsize 
{\large Ben Lillie and JoAnne L. Hewett} \\
\vskip .3cm
Stanford Linear Accelerator Center \\
Stanford CA 94309, USA\\
\vskip .3cm

\end{center}

\begin{abstract} 

We examine the contributions to rare processes that arise in
models where the Standard Model fermions are localized at
distinct points in compact extra dimensions.
Tree-level flavor changing neutral current interactions for
the Kaluza-Klein (KK) gauge field excitations are
induced in such models, and hence strong constraints are
thought to exist on the size of the additional dimensions.
We find a general parameterization of the model
which does not depend on any specific fermion
geography and show that typical values of the parameters
can reproduce the fermion hierarchy pattern.
Using this parameterization, we reexamine the contributions
to neutral meson mixing, rare meson decays, and single
top-quark production in $e^+e^-$ collisions.  We find that it is possible to
evade the stringent bounds for natural regions of the
parameters, while retaining finite separations between the
fermion fields and without introducing a new hierarchy.  The resulting
limits on the size of the compact dimension can be as low as $~$TeV$^{-1}$.
\end{abstract} 

\renewcommand{\thefootnote}{\arabic{footnote}} \end{titlepage} 

\section{Introduction}

In recent years there has been much interest in the possibility that there
may exist compact extra dimensions with sizes far above the Planck
length. 
In particular, the possibility of $\tev^{-1}$-sized extra dimension arises
in braneworld theories
\cite{Antoniadis:1990ew,Lykken:1996fj,Witten:1996mz,Horava:1996ma,Horava:1996qa,Caceres:1997is}.
By themselves, they do not allow for a reformulation of the hierarchy
problem, but they may be incorporated into a larger structure in which
this problem is solved, such as the case of large extra dimensions
\cite{Arkani-Hamed:1998rs,Antoniadis:1998ig,Arkani-Hamed:1998nn}. In the
scenario with $\tev^{-1}$\ extra dimensions, the Standard Model (SM) fields
are phenomenologically allowed to propagate in the bulk.
These models are hence subject to stronger experimental constraints and
have distinct experimental signatures from the case where gravity alone is
in the bulk.

There are many possibilities for how to place the Standard Model fields in
the $\tev^{-1}$ bulk. In the universal extra dimensions scenario all fields see the
extra dimensions, giving rise to a conserved parity that relaxes direct
production and precision electroweak constraints, and may provide a dark
matter candidate
\cite{Appelquist:2000nn,Servant:2002aq,Servant:2002hb,Cheng:2002ej}. 
The effects of universal extra dimensions in rare processes
have been considered in \cite{Buras:2003mk,Buras:2002ej,Agashe:2001xt,Chakraverty:2002qk}.
It is also
possible to localize the fermions without localizing the bosons, which
allows for the gauge fields to propagate freely throughout the bulk.
More recently it was noticed by Arkani-Hamed and Schmaltz (AS) that one
could localize different fermion species at different points in the
$\tev^{-1}$\ extra
dimensions \cite{Arkani-Hamed:1999dc}. These ``split fermion'' models naturally suppress many
dangerous operators, particularly those inducing proton decay. They also
can naturally generate large Yukawa hierarchies; and it has been shown by
multiple authors that there exist models which can generate the correct
spectrum of fermion masses, as well as the correct magnitudes for CKM
matrix elements \cite{Grossman:2002pb,Mirabelli:1999ks,Chang:2002ww,Branco:2000rb,Kaplan:2001ga,DelAguila:2001pu}.
The most stringent generic limits in this case arise from precision electroweak
measurements, which place the compactification radius at $R \lesssim 2-4 \tev^{-1}$
\cite{Rizzo:1999br,Masip:1999mk,Marciano:1999ih,Hewett:2002hv}.
The specific fermion locations can be probed in high energy
collisions, and at very large energies, cross sections will
rapidly vanish since split fermions will completely miss
each other in the extra
dimensions.\cite{Arkani-Hamed:1999za,Rizzo:2001cy} 

This makes the split fermion scenario an attractive possibility for the origin
of the Yukawa hierarchy. However, split fermions (like most models of the
fermion spectrum) are also capable of generating large flavor changing
neutral currents (FCNC). The magnitude of these currents in the neutral meson
sector has been estimated by several
groups, and apparently generate strong
constraints\cite{Chang:2002ww,Delgado:1999sv,Abel:2003fk}.
In this paper we reexamine these computations to derive more model
independent constraints on split fermion models arising from FCNC and show
that it is possible to evade the stringent bounds for natural regions of
the parameters.

This paper is organized as follows. In Section 2 we set up the split fermion 
scenario in as much generality as possible and give
statistical arguments to demonstrate that they can account for the observed fermion
spectrum. We then describe how FCNC are generated in this scenario. In
section 3 we calculate the effects on
neutral meson oscillation. Section 4 presents the effects on rare $B$
decays, and single top production in $e^+e^-$ collisions, and Section 5
concludes.

\section{The Model of Split Fermions}

Here, we construct a very general model that characterizes the
effects of separating the Standard Model fermions in an extra
dimension. We start by examining the original model considered by AS.

In the AS model there is one extra dimension, which is taken to be
flat. It is possible that this extra dimension is actually a ``brane''
with a finite width
embedded in some other extra-dimensional scenario. 
For this reason $\tev^{-1}$\ dimensions are often called ``fat
branes'', but they need not be tied to other models.
Note that if the brane is not a string theory object, but arises from
some field theory mechanism, then it {\it necessarily} has finite extent
in the extra dimensions. This makes the study of fat branes essential to
building realistic field-theoretic models of extra dimensional scenarios. 

In this model the Standard Model fields are localized to the brane.
Note that the word {\it brane} here refers to any mechanism for achieving this
localization. It may or may not be the same as the branes encountered in
string theory.
Initially, all fields are allowed to propagate in the entire
dimension. In addition to the fields present in the Standard Model, there
is a real scalar field which couples to the fermions, but not
to the gauge bosons or the Higgs.

If the scalar has a $Z_2$-symmetric potential, then it can
develop a stable solution which tunnels from one of the vacua to the
other, called a kink solution. 
A mechanism for localizing fermions to a thin but finite width region inside
a domain wall has been known for some time \cite{Jackiw:1976fn}. There it was
noted that in 1+1 dimensions a massless fermion with a Yukawa coupling to a scalar
field that has a kink-profile vacuum expectation will develop a zero
mode, with a Gaussian profile centered at the location of
the kink. This can be trivially extended to more dimensions 
by considering a domain wall instead of a soliton
and making all zero modes constant in the
transverse directions. Note that a five-dimensional fermion field contains two
four-dimensional fermions, one of each chirality. If the extra dimension
is infinite, then the zero-mode of only one chirality is
normalizable. If the extra dimension is finite then something else is
needed to produce chirality. A standard procedure is to compactify the
dimension on an $S^1/Z_2$ orbifold, which projects out the unwanted
chirality. A nice side-effect of this is to render the kink absolutely
stable.\footnote{It is interesting to think that if one invokes a mechanism
to localize the gauge bosons, as in \cite{Dvali:2000rx}, then one could
have a fat brane residing in an infinite dimension.}

In contrast to the fermion sector, the gauge bosons are free to propagate
throughout the extra dimension. Since
the dimension is compact, and flat, the
mass spectrum of the Kaluza-Klein gauge states is linear with $M_n^2 =
n^2/R^2$, and the orbifold boundary conditions project out the odd
solutions, so the wavefunctions along the fifth dimension, $y$ where $0
\le y \le R$, are
\begin{gather}
A^{(n)\mu}(x,y) = \sqrt{\frac{2}{R}}\cos(\frac{n\pi
y}{R})A^{\mu}(x)\label{eq:gaugewf} 
\hspace{1cm}(n\ge 1),
\end{gather}
where $R$ is the size of the extra dimension. Putting all this together
allows investigation of brane world
models where there is a single extra-dimension of roughly inverse $\tev$
size with fermions localized in the center and gauge bosons propagating
though the entirety.

A more interesting picture can be obtained by thinking about the fermion
localization mechanism. There is a simple heuristic for why this should
occur. The fermion is Yukawa coupled to a scalar field which develops a
non-zero VEV. The ordinary fermion Higgs phenomena should then give the
fermion a mass. However, the VEV is position dependent and in particular
there is a place where it is zero (the center of the kink). So the fermion
has a position dependent mass, which is somewhere zero. Thus, the fermion
is easiest to excite near the zero mass, and so most of the probability
for the lowest lying state (the zero mode) will live near the center of
the kink.

Given that heuristic, it should be reasonable that if the 5D fermion has a
mass $M$, then
the center of the Gaussian moves to $y = M/2\mu^2$, where $\mu$ is the
slope of the kink profile, and $v$ is the scale of the VEV. Indeed, it
turns out that this is the case, as was first noted by Arkani-Hamed and
Schmaltz \cite{Arkani-Hamed:1999dc}. This allows different fermion fields to be
localized at different points in the extra dimension. 
To see why this is desirable, consider an operator, $\cal O$, that
involves fermions separated by a distance $d$.
The effective 4D coupling in the dimensionally reduced theory is
proportional to the integral over the extra dimensions of the wavefunctions of
all fields appearing in $\cal O$.
Since the fermion wavefunctions are Gaussian, this gives a
suppression proportional to $e^{-a\mu^2 d^2}$, where $a$ depends on
the operator being considered. 
This has been shown to be very effective at suppressing
dangerous higher-dimensional operators, such as proton
decay. Additionally, the fact that exponentially different couplings can
result from linear separations provides a natural means of explaining the
fermion mass hierarchy. Lighter fermions have greater separation between
their left and right handed components. In this way Arkani-Hamed and
Schmaltz proposed a theory to explain the Yukawa
hierarchy without invoking new symmetries, and which is safe from proton
decay. Several authors have proposed 
specific ``geographies'' that do indeed reproduce the correct fermion
masses, as well as the CKM
parameters \cite{Grossman:2002pb,Mirabelli:1999ks,Chang:2002ww}.

\begin{figure}[thbp]
\centerline{
\includegraphics[bb = 140 583 265 682,width=7cm,angle=0]{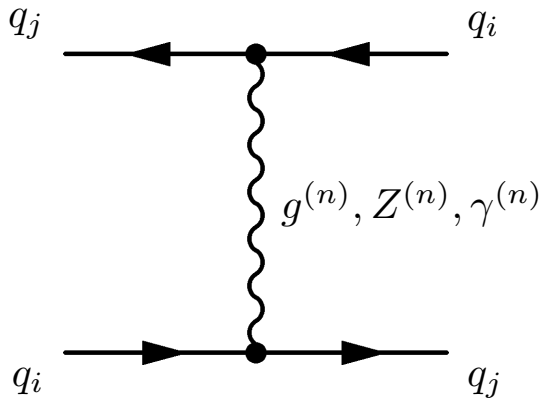}}
\caption{\small
Feynman diagram for the tree-level KK gauge exchange mediating neutral
meson oscillation.
}
\label{fig:fcncgraph}
\end{figure}

There are, however, other potentially dangerous effects of the
fermion separation which are not suppressed by this mechanism. 
The gauge bosons will have a Kaluza-Klein (KK) tower of
states. The zero modes, which are flat in the extra dimensions, correspond to the SM
gauge fields, and have the correct couplings to the fermion zero modes. On
the other hand, the
excited states have cosine profiles, as given in
Eq. (\ref{eq:gaugewf}).\footnote{In general they are
exponentials, $e^{iny}$, but the orbifolding projects out the odd modes.}
The coupling strength of these modes to the fermions are scaled by an
integral over the overlap of the fermion and gauge wavefunctions. However,
since the height of the boson wavefunction will be different at the
locations of the different fermions, there will be non-universal couplings of a single gauge KK-state to different fermion
species. This leads to the possibility of flavor changing interactions,
including tree-level neutral currents, for the KK-modes of the
$\gamma$, $Z$, and gluon, as illustrated in Fig \ref{fig:fcncgraph}. One then expects large effects to
come from the tree-level contributions of the KK gluon states to FCNC
processes, in particular to neutral meson
oscillation. Calculation of these effects can put limits on the size of
the extra dimension. Also, note that while this discussion was motivated by the kink model, these issues will
be relevant to any model with split fermions.
This is an example of the general principle that any
attempt to explain the Yukawa hierarchy will necessarily treat flavors
differently, and will tend to generate large flavor-changing effects.

In practice, geography independent constraints have been difficult to
obtain due to the
large number of parameters in the model.
These are, $R$, $\sigma$, the
width of the fermion wavefunctions (which is $1/\mu$ in the kink model),
and $(d^n - d)$ positions, where $d$ is the number of extra dimensions and
$n$ is the number of independent fermion fields.
Previous discussions \cite{Chang:2002ww,Delgado:1999sv,Abel:2003fk} have
put constraints on $R$ only by first
obtaining a single set of positions that reproduce the Yukawa couplings of
the Standard Model, and calculating the flavor-changing effects in that
particular geography. However, one would like a more model-independent way of
understanding the magnitude of flavor effects in this class of models.

To accomplish this we consider the problem of FCNC in split fermion models in
as much generality as possible. A specific, realistic model exists in
string theory \cite{Abel:2003fk}, as well as the field theory example just
presented. In summary, we abstract from these the following points:

\begin{enumerate}

\item There exist one or more extra dimensions, compactified with a radius
$R$.\footnote{In one dimension the compactification is $S^1/Z_2$. In more
dimensions we take the compactification to be flat, and orbifolded in such
a way that it looks like a simple product of single dimensions.}

\item Each fermion field, $\psi_i$ has a chiral zero mode that is
localized near the center of the dimension at $y_i$, with Gaussian
profiles $\psi \sim e^{-(y-y_i)^2/\sigma^2}$, where the width $\sigma \ll R$. If there is more than one
extra dimension they are taken to be isotropic in those dimensions $\psi
\sim e^{-(\vec{y}-\vec{y_i})^2/\sigma^2}$.

\item The gauge bosons are free to propagate in the entire ``fat brane''
part of the extra dimension. (There may be a larger bulk accessible to
gravity.)

\item The boundary conditions for the bosons are taken to be such that the
wavefunction for the $n$-th KK-mode is $A \sim \cos(n \pi y/R)$. Note that
these are generally the same conditions that allow chiral zero modes for the
fermions.

\item The field content (gauge group, number and charge of matter fields)
is identical to the Standard Model, plus whatever fields are necessary to
localize the fermions.

\end{enumerate}
These assumptions generate an effective four-dimensional Lagrangian that
reduces to the Standard Model at low energies. The new features present
are the propagating gauge KK-modes, their couplings, and the fact that the
Higgs Yukawa couplings are determined by the fermion locations.

We now construct the interaction Lagrangian for this scenario, focusing on
the quark sector in this paper. An analogous treatment of the leptonic
sector can be performed.
Note that there are excited states of the fermion fields in addition to
the KK
boson states. However, since the fermions are localized with a width smaller
than R, the
scale of the fermion excitations will be significantly higher
than that of the KK gauge states. In addition, the fermion KK modes do not
participate in the processes considered here. We therefore only consider
the fermion zero modes, while we include the complete KK-tower for the bosons.
With one extra dimension the coupling of the $n$-th KK boson to a flavor
localized at the scaled position $\ell = x/R$ is determined by the overlap
of wavefunctions
\begin{gather}
\int_0^1 dy \bar \psi(y) \psi(y) A^{(n)}(y) \simeq
\int_0^1 dy \cos\left(n\pi
y\right)e^{-(y-\ell)^2 R^2/\sigma^2} \simeq \cos\left(n\pi
\ell\right)e^{-n^2 \sigma^2/R^2}.\label{eq:gaugecoupling}
\end{gather}
where y has now been normalized to R. For $\delta$ extra $\tev^{-1}$\
dimensions this generalizes to
\begin{gather}
c^{(\vec n)}(\ell) \equiv 
\left(\prod_{k=1}^{\delta}\cos\left(n_k\pi \ell_k\right)\right)e^{-{\vec n}^2
\sigma^2/R^2}.
\end{gather}
The gauge coupling of the gluons, for instance, can then be written in
flavor space as
\begin{gather}
{\cal L}_{\rm int} = \sqrt{2} g_s G_{(n)\mu}^{A} \left(\bar{\bf d}_L
\gamma^{\mu} T^{A} C^{(n)}_{L}{\bf d}_L
+ \bar {\bf d}_R \gamma^{\mu} T^{A} C^{(n)}_{R}{\bf d}_R\right)
+ ({\bf d} \to {\bf u}) + {\rm h.c.}.
\end{gather}
Here ${\bf d}_{L (R)}$ is the vector of left (right) handed down-type
quarks, $\bar{\bf d} = (d\ s\ b)$, $g_s$ is the SU(3) coupling constant, and $G^{(n)\mu}$ is
the $n$-th KK gluon field. The diagonal matrices $C^{(n)}_{i}$ are the
wavefunction overlaps given by Eq. (\ref{eq:gaugecoupling}). The factor of
$\sqrt{2}$ arises from the rescaling of the gauge kinetic terms to the
canonically normalized value for all $n$.

Now, the Higgs zero mode, which is the Standard Model Higgs, is flat in
the extra dimension, $H^0
\propto 1/R$. Then the Yukawa couplings to the 4D Higgs field are given by
\begin{gather}
R\lambda_5\int_0^1 dy H^0 \bar q_L q_R 
\simeq \lambda_5\int_0^1 dy\ e^{-(y-y_i)^2/\sigma^2} e^{-(y-y_j)^2/\sigma^2}
\simeq \lambda_5 e^{-(y_i - y_j)^2/\sigma^2}.\label{eq:yukawacouplings}
\end{gather}
Here $\lambda_5$ is an overall 5D coupling constant that is fixed to be
${\cal O}(1)$ by the top quark mass. We write the 4D Yukawa couplings to
(for instance) the down-type quarks in the flavor basis as
\begin{gather}
{\cal L}_{\rm Yukawa} = \bar d V_R^{(d)\dagger}M_d V_L^{(d)}d
\end{gather}
Where $V_R^{(d)\dagger}M_d V_L^{(d)}$ is the matrix of Yukawa couplings
with elements given by Eq. (\ref{eq:yukawacouplings}), and
$M_d$ is the diagonal mass matrix.

We can now write the relevant terms of the Lagrangian as
\begin{align}
{\cal L} = 
\bar {\bf d}_L V_R^{(d)\dagger}M_d V_L^{(d)}{\bf d}_R
+\bar {\bf u}_L V_R^{(u)\dagger}M_u V_L^{(u)}{\bf u}_R
+ \frac{g}{\sqrt{2}}W^{(0)}_{\mu} \bar {\bf u}_L \gamma^{\mu} {\bf d}_L\notag\\
+ \sum_{n = 1}^{\infty} \left[ \sqrt{2} g_s G^{(n)A}_{\mu} \left(\bar {\bf d}_L
\gamma^{\mu} T^{A} C^{(n)}_{L}{\bf d}_L
+ \bar {\bf d}_R \gamma^{\mu} T^{A} C^{(n)}_{R}{\bf d}_R\right) + ({\bf d} \to {\bf
u}) \right] + {\rm h.c.}
\end{align}
After the usual transformation to the mass basis, the CKM-matrix is
clearly the product $V^{(u)\dagger}_L V^{(d)}_L$. Note, however, the
presence of non-universal couplings prevents the products $U^{q(n)}_i
\equiv V^{(q)\dagger}_i C^{(n)}_i V^{(q)}_i$ from being trivial, so there
are flavor-changing interactions in the KK-gluon sector. These also occur
in the excited photon and $Z$ couplings. However, those are suppressed
relative to the gluons by a factor of $g/g_s$, so we expect that the
KK-gluons will dominate any process to which they contribute.

Before examining the numerical impact of the tree-level FCNC interaction in
rare processes, it will first be useful to get a handle on how far the
fermions need to be
separated. It has been shown by Grossman and Perez \cite{Grossman:2002pb} that there
exists at least one set of positions that correctly reproduces the
observed fermion spectrum and magnitude of the CKM elements. They found
that, subject to a certain set of naturalness assumptions, there was a
single solution. A different solution was found in \cite{Chang:2002ww} by choosing
different up and down-type Yukawa coupling constants in the 5D
theory. Typical separations in these solutions are from $1-20$ units of
the fermion width. In what follows, we parameterize the separation between 2
fermions in units of the width, i.e. $\Delta y =y_i - y_j =
\alpha_{ij}\sigma$, and treat $\alpha_{ij} $ as phenomenological parameters. In
addition, we find it useful to define $\rho = \sigma/R$.

\begin{figure}[t]
\centerline{
\includegraphics[width=18cm,angle=0]{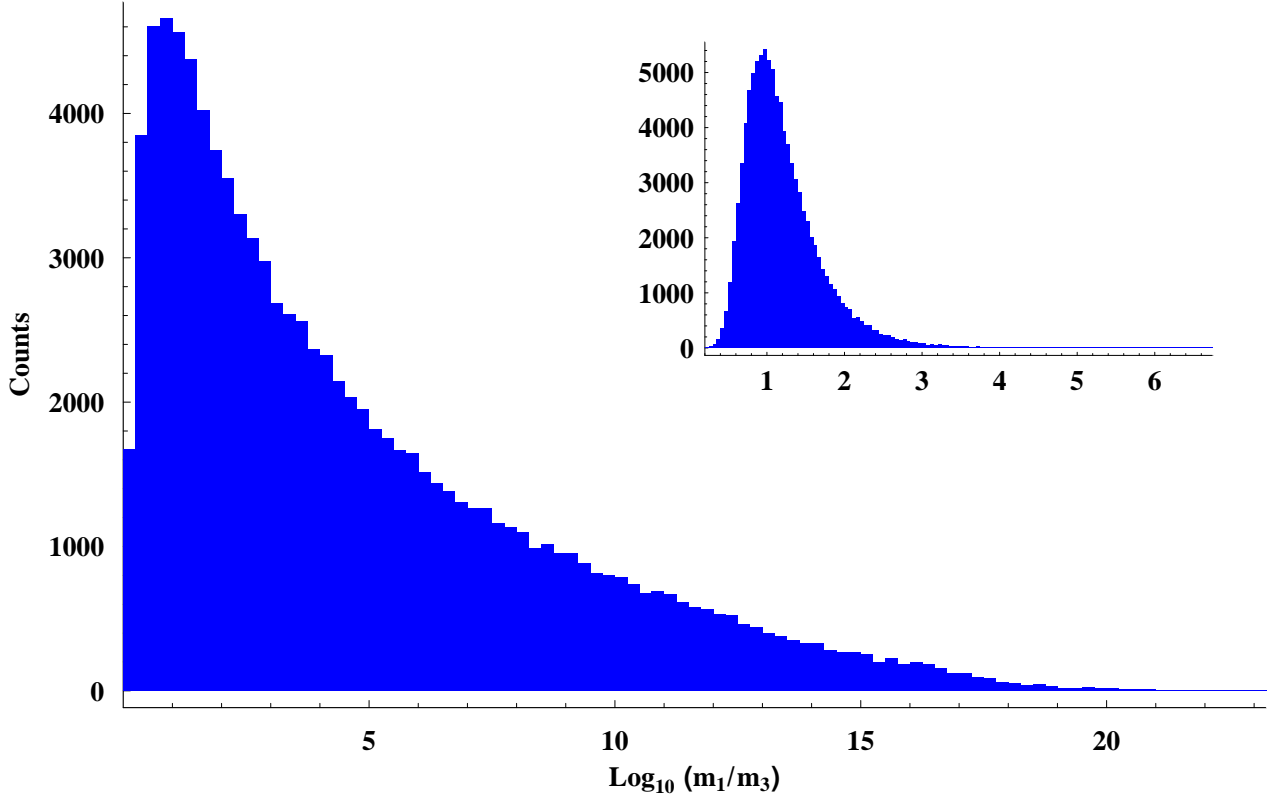}
}
\caption{\small
Histograms of $10^5$ random trials of the size of the fermion hierarchy,
$\log(m_1/m_3)$, where
$m_1$ is the largest and $m_3$ the smallest mass for a single Yukawa
matrix. Main graph: In the split fermion model where the positions are randomly
drawn on the interval $[0,15]$ in units of the fermion width. Inset: In a
``null hypothesis'' where the Yukawa matrix elements are randomly drawn
from the interval $[0,1]$.
}
\label{fig:data15}
%\end{figure}

%\begin{figure}[htbp]
%\caption{\small
%Caption: Null Hypothesis
%}
%\label{fig:nulldat}
\end{figure}

As a counterpoint to the studies in \cite{Grossman:2002pb} and
\cite{Chang:2002ww} we have performed a simple Monte-Carlo
analysis in an attempt to see how large of a hierarchy is generated
naturally for fermions randomly distributed on an interval. To do this we
randomly draw fermion positions from a distribution flat on the interval
$[0,\alpha_{\rm max}]$, and use these to compute the Yukawa matrices from
Eq. (\ref{eq:yukawacouplings}). We then compute the singular values of
these matrices, which are the fermion masses. We can get a sense of the
hierarchy by taking a particular Yukawa matrix (say the up-type) and
finding the ratio of the largest to smallest singular value. In Fig
\ref{fig:data15} we show a histogram of the log of this ratio for
$\alpha_{\rm max} = 15$. For comparison we have computed the same value
for a ``null hypothesis'' where instead of the split fermion scenario, the
entries of the Yukawa matrices 
themselves are
drawn directly from a distribution flat on the interval $[0,1]$. As
expected, the case of split fermions clearly generates a much larger
hierarchy. What is surprising is that one needs to set $\alpha_{\rm max}
\approx 10-15$ before a hierarchy of six orders of magnitude becomes
common, while in \cite{Arkani-Hamed:1999dc} it was claimed
this hierarchy could result from $\alpha_{\rm max} \approx 5$. The
discrepancy is due to the fact that, while a separation of $\alpha = 5$
will indeed generate a matrix element of order $10^{-6}$, the singular
values (which are the actual masses) of a full Yukawa matrix with
separations no larger than 5 will tend to be too large. We note that the
full fermion spectrum can be generated by $\rho$ as large 
as $1/15$, i.e. without introducing a new large hierarchy between the
compactification and fermion localization scales.
Also note that $\alpha_{\rm max}$ represents the part of the extra dimension in
which the fermions can be localized and need not be the same as $1/\rho$,
which is the size of the dimension through which the gauge bosons can
propagate.

%\begin{figure}[thbp]

\section{Constraints from Neutral Meson Oscillation}

Significant effects from the flavor-changing gluonic couplings should show up
in neutral meson oscillations. We start by examining the effects on Kaon
mixing. The $\Delta S = 2$ effective Lagrangian
from the single KK-gluon exchange depicted in Fig. \ref{fig:fcncgraph} is
\begin{align}
&{\cal L}_{\rm eff}^{\Delta S = 2}  = \frac{2}{3}
g_s^2\sum_{\vec{n}=1}^{\infty} 
\frac{1}{M_n^{2}}
\sum_{i,j = L,R} U_{i(sd)}^{d(n)*}U_{j(ds)}^{d(n)}
\bar d_i \gamma^{\mu} s_i \bar d_j
\gamma_{\mu}s_j \notag\\& =
\frac{2}{3} g_s^2\left(\vphantom{\frac{1}{2}}\right.
\sum_{i,j=L,R}
V^{d}_{i(11)}V^{d*}_{i(12)}V^{d*}_{j(11)}V^{d}_{j(12)}\notag\\ 
&\times \left.\vphantom{\frac{1}{2}}\times \sum_{n=1}^{\infty}
\frac{(\cos(n \pi x_{d_i}) - \cos(n \pi x_{s_i}))(\cos(n \pi x_{d_j}) -
\cos(n \pi x_{s_j}))}{M_n^{2}}e^{-\rho^2 n^2}
\right).\label{eq:efflagrangian}
\end{align}
Here, the $x_i$ are the positions of the $d$ quark fields, $y_i$ of the $s$
quark, and we have used the unitarity of the $V$'s (here, we have
approximated this with $2\times 2$ unitarity, we
discuss the third-generation effects below).
We are especially interested in the form of the sum over the KK-modes. While
KK-sums are usually divergent in more than one extra dimension and require
a cutoff, ours contains a natural cutoff arising from the finite width of
the fermion zero mode, and hence is convergent for a number of extra
dimensions $\delta \ge 1$.
This is simple to understand physically. The
cutoff sets in when the
wavelength of the KK-mode is of order the fermion width, which occurs at
$R/n_{\rm max} = \sigma$. At higher momenta
the wavefunction of the boson is oscillating many times within the
fermion allowing it to resolve the fermion's wavefunction, and
exponentially decouples. The fact that this cutoff arises naturally
in the field theory model is an attractive feature of that particular
mechanism for fermion localization.

In one additional dimension the sum converges even without the exponential
suppression. In this case it is insensitive to the value of $\rho$ and can
be computed analytically by ignoring the exponential factor. To do this we
need to evaluate
\begin{gather}
F(x,y) \equiv \sum_{n=1}^\infty \frac{(\cos(n\pi x)-\cos(n\pi y))^2}{n^2},
\label{eq:fnotsummed}
\end{gather}
and
\begin{gather}
G(x,y) \equiv \sum_{n=1}^\infty \frac{(\cos(n\pi x_1)-\cos(n\pi
y_1))(\cos(n\pi x_2)-\cos(n\pi y_2))}{n^2}.
\end{gather}
The computations for these sums are presented in the appendix. The final
result is
\begin{gather}
F(x,y) =
\frac{\pi^2}{2}|x-y|,\label{eq:f}
\end{gather}
and
\begin{gather}
G(x_1,y_1,x_2,y_2) = 
\frac{\pi^2}{2}
(|x_1-x_2| + |y_1-y_2| - |x_1 - y_2| - |x_2 - y_1|).
\end{gather}
This tells us that the flavor changing effects depend, as expected, on any
nonzero separation between fermion fields.

The hadronic matrix elements for the gluonic contributions to Kaon mixing
are given by (computed in the vacuum insertion
approximation)\cite{Gabbiani:1996hi}:
\begin{align}
\langle \bar K^0|\bar d_L\gamma^\mu s_L \bar d_L \gamma_\mu
s_L|K^0\rangle
& = \frac{1}{3}f_K^2 m_k\\
\langle \bar K^0|\bar d_L\gamma^\mu s_L \bar d_R \gamma_\mu
s_R|K^0\rangle
& = f_K^2 m_k\left(
\frac{1}{12} + \frac{1}{4}\left(\frac{m_K^2}{m_d^2 + m_s^2}\right)
\right)\notag
\end{align}
as well as those with $(L \leftrightarrow R)$, which have the same
evaluation.
Written out in full, the contribution to $\Delta m_K$ is then
\begin{align}
\Delta m_K = 
&\Re\langle \bar K^0|{\cal L}^{\Delta S =2}|K^0\rangle
\notag\\ 
= \frac{2}{3}g_s^2 R^2
&\left(\vphantom{\frac{1}{1}}\right.
|V^{d}_{L\ 11}V^{d*}_{L\ 12}|^2
F(x_{d_L},x_{s_L})
\langle \bar K^0|\bar d_L\gamma^\mu s_L \bar d_L \gamma_\mu
s_L|K^0\rangle\notag\\
&+|V^{d}_{R\ 11}V^{d*}_{R\ 12}|^2
F(x_{s_R},x_{s_R})
\langle \bar K^0|\bar d_R\gamma^\mu s_R \bar d_R \gamma_\mu
s_R|K^0\rangle\label{eq:ds2}\\
&+(V^{d}_{L\ 11}V^{d*}_{L\ 12}V^{d*}_{R\ 11}V^{d}_{R\ 12})
G(x_{d_L},x_{s_L},x_{d_R},x_{s_R})
\langle \bar K^0|\bar d_L\gamma^\mu s_L \bar d_R \gamma_\mu
s_R|K^0\rangle\notag\\
&+(V^{d}_{R\ 11}V^{*d}_{R\ 12}V^{*d}_{L\ 11}V^{d}_{L\ 12})
G(x_{d_L},x_{s_L},x_{d_R},x_{s_R})
\langle \bar K^0|\bar d_R\gamma^\mu s_R \bar d_L \gamma_\mu
s_L|K^0\rangle
\left.\vphantom{\frac{1}{1}}\right),\notag
\end{align}
where $x_{d_{L,R}}$ are the positions of the $d$ field, and the $x_{s_{L,R}}$ of
the $s$. 
Note that all possible separations (between quark fields) are present, but some
enter with different signs. In principle then, the gluonic contribution
could be made
small for any values of $R$ and $\rho$ by placing the quarks at
appropriate places. However, the terms involving only right or
left handed fields occur with the same sign. So, to achieve significant
reduction, cancellations must occur between those terms and the terms which
involve both chiralities. This involves tuning the quark positions to the
values of different hadronic matrix elements, which introduces a
fine-tuning problem. Otherwise, it would imply that the UV physics
that localizes the quarks has information about the IR behavior of QCD!
We can therefore expect that cancellations
will be ${\cal O}(1)$ at most. This is seen clearly in the Monte Carlo
trials, where the random positions have no relation to the hadronic matrix
elements and no significant cancellation occurs.

In light of this we can explore the magnitude of the flavor effects just
by looking at a 
single term in (\ref{eq:ds2}); for convenience we choose the first. We
can then describe the contributions with only three parameters: the radius
$R$, the scale
ratio $\rho$, and the separation between one pair of fermions,
$\alpha$. The sum over KK-modes is then calculable in terms of these
parameters. Since $R$ enters only in the mass in the KK propagator, we can
write the contribution in the simple form
\begin{gather}
\Delta m_K = \frac{2}{9}g_s^2 f_K^2 m_K R^2 V_{\rm ds,ds}^4 F_{\rho}(\rho\alpha)
\end{gather}
where  $V_{ds,ds}^4$ stands for the appropriate product of 4 elements of the
$V^{u,d}_{L,R}$ matrices, two at each vertex, and, as
shown in (\ref{eq:f}), $F(x,y)$ only depends on the difference of it's
arguments, so we can write it as a function of only a single variable,
given by the product
$\rho\alpha$. The subscript on F reminds us that in two dimensions or more
$F$ also depends on $\rho$ directly as the cutoff parameter, in which case
it must be computed numerically. If we demand that this contribution to
$\Delta m_K$ be no
larger than the measured value (a conservative assumption from the point of
view of constraining the model) we get
\begin{gather}
\frac{1}{R} \ge \beta_K \sqrt{V_{ds,ds}^4 F_{\rho}(\rho\alpha)},\label{eq:defbeta}
\end{gather}
where $\beta_K$ is a coefficient of dimension 1, which depends on the
meson parameters. This expression immediately
generalizes to other neutral meson systems by using the appropriate
coefficient $\beta$, and the appropriate matrix elements of $V$. Table
\ref{table:sum} shows the values of $\beta$ for cases of interest, along
with representative values of $F_{\rho}(\rho\alpha)$.

\begin{figure}[t]
\centerline{
\includegraphics[width=12cm,angle=0]{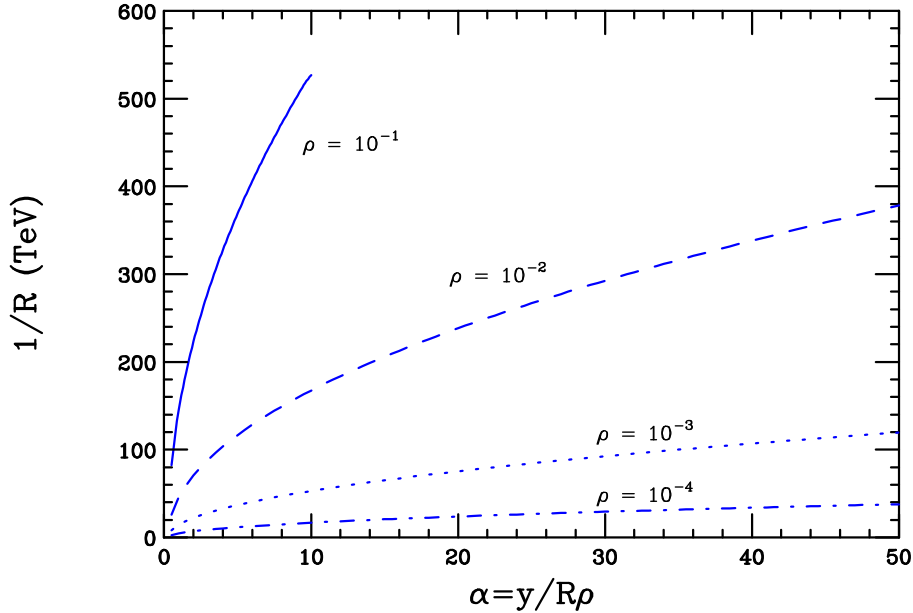}}
\caption{\small
The behavior of the constraint on $1/R$ with $\alpha$ for
various values of $\rho$. The area below the curves is excluded. Note that
the size of the additional dimension in units of the
fermion width is $1/\rho$, so the curve for $\rho = 1/10$ ends at a
maximal separation of $\alpha = 10$.
}
\label{1kaonfig}
\end{figure}

\begin{figure}[t]
\centerline{
\includegraphics[width=12cm,angle=0]{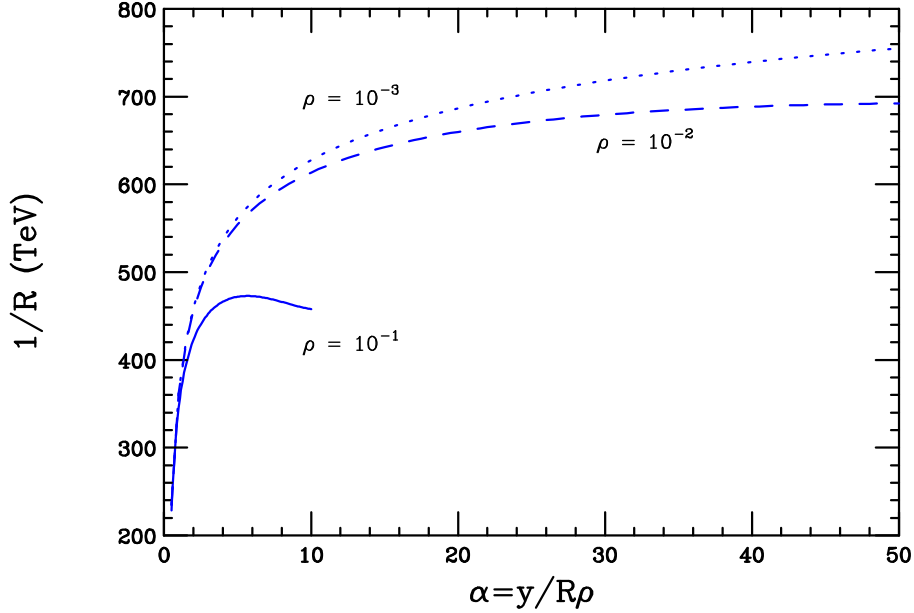}}
\caption{\small
Constraints on $1/R$ from Kaon mixing for two extra dimensions; the area
below the curves is excluded.
}
\label{2kaonfig}
\end{figure}

The resulting constraints are shown in Fig. \ref{1kaonfig} and
\ref{2kaonfig}, for 1 and 2 extra dimensions respectively, using the value
of $\beta$ 
and $V_{ds,ds}^4$ appropriate for the Kaon sector, and assuming that the $V$ are
CKM-like
in magnitude (we discuss that assumption in detail below). 
There are two
features of note. First, with one extra dimension the constraint
is a simple square-root function, as can be seen from Eq. (\ref{eq:f}). 
This means
that the flavor-changing effects can be made arbitrarily small by reducing
$\rho$. That is, by increasing the hierarchy between fermion and boson
scales. Second, in two dimensions the effect seems to be roughly constant
in $\rho$, and flattens off at large $\alpha$. We know that the sum over
the KK states is
diverging logarithmically before it
reaches the cutoff, and so it should be getting larger as
$\rho$ decreases. However, shrinking $\rho$ brings the fermions closer together
making the flavor effects smaller. In two dimensions these two effects are
seen to roughly cancel. In three or more dimensions, the divergence of the
sum wins
completely, and the bounds on $R^{-1}$ are huge, effectively removing these
cases from consideration as realistic models.

\begin{figure}[thbp]
\centerline{
\includegraphics[width=12cm,angle=0]{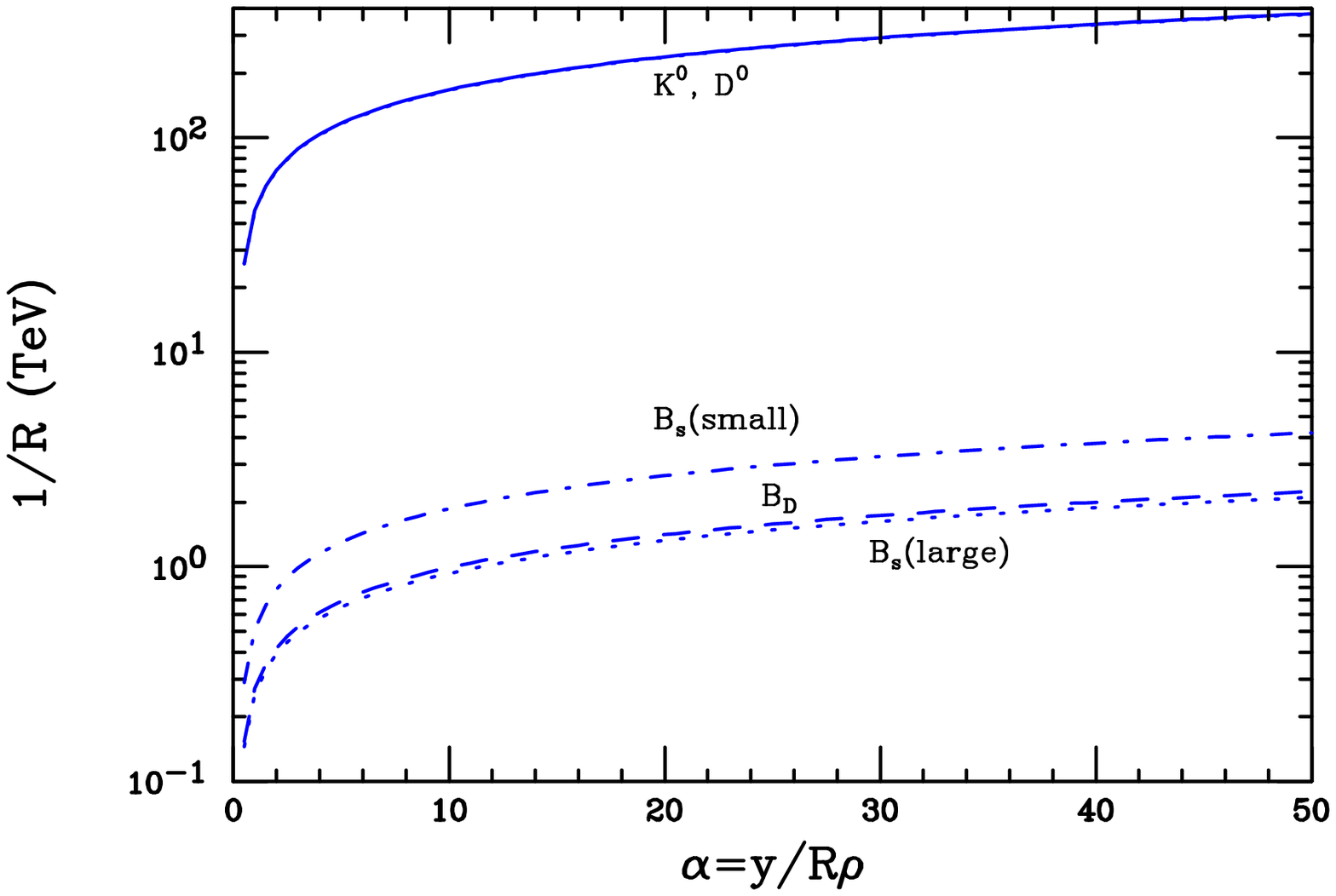}}
\caption{\small
Constraints from all species of neutral meson mixings for one extra
dimension, taking $\rho =
1/100$. See text for a description of the experimental values used. Note
that the $K^0$ and $D^0$ results are separate lines that overlap due to a
numerical coincidence.}
\label{1allfig}
\end{figure}

\begin{figure}[thbp]
\centerline{
\includegraphics[width=12cm,angle=0]{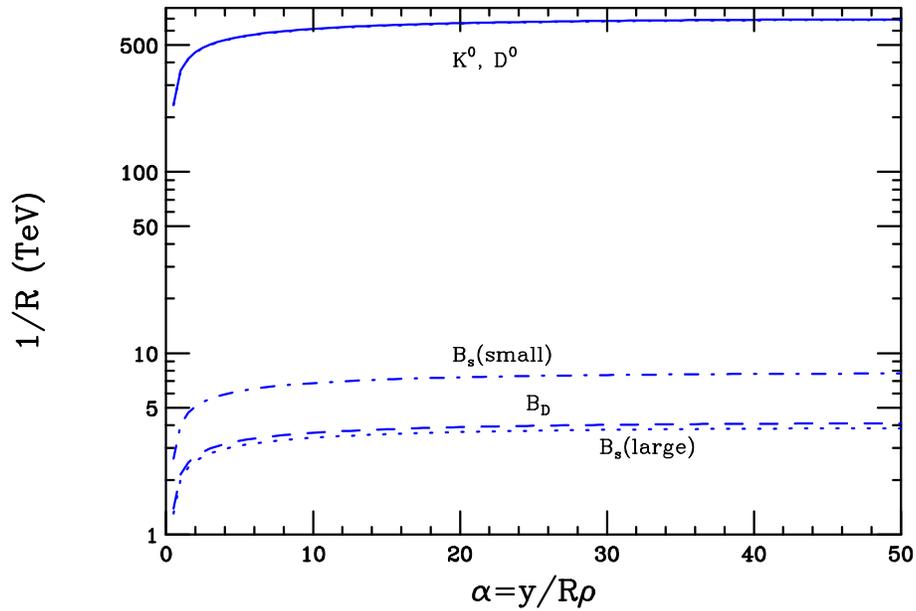}}
\caption{\small
Neutral meson constraints for two extra dimensions. The area below the
curves is excluded.
}
\label{2allfig}
\end{figure}

In Figs. \ref{1allfig} and \ref{2allfig} we display our results for all
meson mass differences, taking $\rho = 1/100$.
For mixing in the Kaon sector and $B_d^0$ sector, the bound is set by demanding that the
new physics produce an effect which is no larger than the observed value. For
$D^0$ mixing, the effect is restricted to lie below the current experimental
bound. There is no experimental upper bound on $B_s^0$ mixing, so we assume
two values, one the size
expected in the Standard Model, the other about 4 times larger,
corresponding to the curves labeled small and large, respectively.
Note that the most stringent constraints
come from mixings involving the first and second generation.

\begin{table}
\centerline{
\begin{tabular}{|l|l|l|l|l|l|}
\hline
$\rho$ & $\alpha=1$ & $\alpha=5$ & $\alpha=10$ & $\alpha=20$& $\alpha=50$ \\
\hline\hline
1D & & & & & \\
\hline
$10^{-1}$ &1.15&2.33&4.76&NA&NA\\
$10^{-2}$ &0.036&0.23&0.48&1.00&2.45\\
$10^{-3}$ &0.0012&0.023&0.048&0.097&0.25\\
\hline
2D & & & & & \\
\hline
$10^{-1}$ &2.05&3.82&3.59&NA&NA\\
$10^{-2}$ &2.25&5.27&6.46&7.47&8.22\\
$10^{-3}$ &2.27&5.41&6.67&8.08&9.78\\
\hline
\end{tabular}
\hspace{1cm}
\begin{tabular}{|l|l|}
\hline
Meson & $\beta$(\mev)\\
\hline
$K^0$ &1125.86\\
$B^0_d$ & 478.01\\
$B^0_s$ & 67.4346\\
$D^0$ & 1124.23\\
\hline
\end{tabular}
}
\caption{\label{table:sum} Left: Representative values of the sum
$F_{\rho}(\rho\alpha)$ for one and two extra dimensions. Right:
Multiplicative $\beta$ factors for mass splittings of the
neutral mesons; $1/R \ge \beta\sqrt{V^4 F_{\rho}(\rho\alpha)}$.}
\end{table}

This pattern suggests a loop-hole in the otherwise stringent constraints.
Namely, the $V$ matrices need
not be CKM-like. Since the CKM matrix is the product
$V_L^{(u)\dagger}V_L^{(d)}$, the observed CKM hierarchical structure could
result from a
completely different structure at the level of the $V^{i}$. If there is
small first to second generation mixing the Kaon and $D^0$ constraints
will be relaxed, and all constraints would then be of order a few \tev,
even for large values of $\rho$.

\begin{figure}[thbp]
\centerline{
\includegraphics[width=12cm,angle=0]{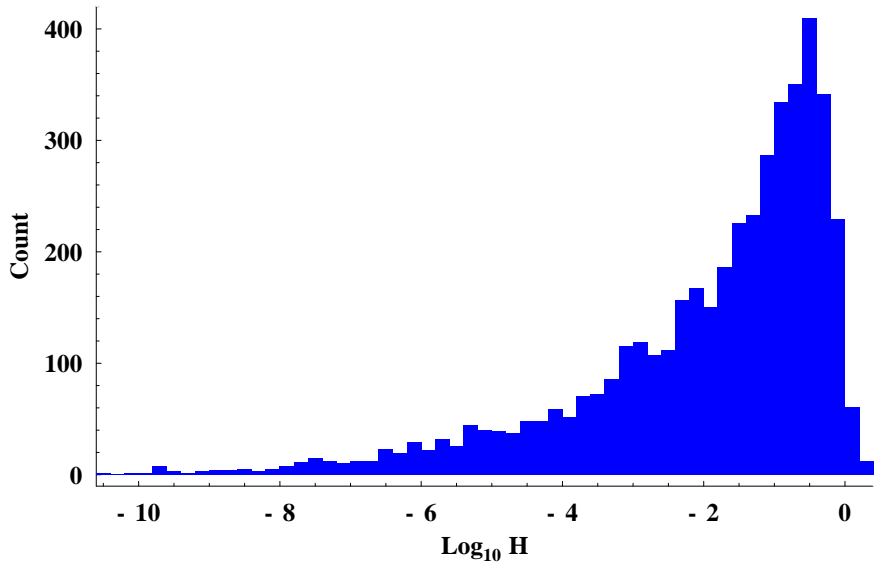}
}
\caption{\small
Histogram of values of the factor $H = \sqrt{\sum V^4 F(x,y)}$ summed over the
appropriate fermion positions and $V$ matrix elements.
}
\label{fig:unitarity}
\end{figure}

However, in the previous calculation we ignored the third generation when
imposing the unitarity condition in Eq. (\ref{eq:efflagrangian}). 
Transitions between two 4D mass eigenstates will involve all three
generations in the localization (flavor) basis. In this case the matrices
$U^{(n)}_{i}$ will contain the positions of all three generations of
quarks, and the unitarity conditions on the $V$ matrices will be changed. 
For instance, for the term in Eq. (\ref{eq:ds2}) with both left-handed chiralities (and dropping
the $L$ index), in place of 
$|V^{d}_{L\ 11}V^{d*}_{L\ 12}|^2
F(x_d,x_s)$ we should have
\begin{align}
|V_{11}|^2 V_{12} V^{*}_{13} F(x_d,x_s)
+ V^{*}_{11} V_{12} V^{*}_{32} V_{31} G(x_d,x_s,x_b,x_s)\notag\\
+ V^{*}_{31} V_{32} V^{*}_{13} V_{11} G(x_b,x_s,x_d,x_s)
+ |V_{31}|^2 |V_{32}|^2 F(x_b,x_s)
\end{align}
These additional terms (including the ones not displayed above corresponding to
right and mixed chiralities) will insure that in any mass splitting
observable many mixing angles and fermion separations will enter. It then
becomes non-trivial to reduce $\Delta m_K$ by adjusting mixings
alone. However, it is still possible to reduce the Kaon and $D^0$ constraints by noticing that
the new terms contain fewer diagonal elements. Hence, if the weak and mass
eigenbases are not too badly misaligned, i.e. the $V^{i}$ have large
diagonal elements and smaller off-diagonal elements, then the strongest
constraints may be relaxed somewhat.
To get a better sense of what is typically possible, we again run Monte Carlo
simulations. From these we learn that a typical suppression factor is
$10^{-1}$ for the factors multiplying $\beta$ in Eq. (\ref{eq:defbeta}), $10^{-2}$ is not uncommon,
and $10^{-4}$ is obtainable, but rare, even for the fairly large value
$\alpha_{\rm max} = 15$. This is illustrated in Fig. \ref{fig:unitarity}.

All of these considerations together show that once the parameter space is
thoroughly explored, it is possible to evade the large constraints from
meson mixing for natural regions of the parameters.

\section{Rare Decays}

We also consider processes involving only a single flavor-changing
vertex. The best examples of this type which receive contributions from KK
gluon exchange are rare $B$ decays, such as $B
\to \psi K_S$ and $B \rightarrow \phi K_S$. The most interesting aspects
of these decays are, of course, their associated CP-violating asymmetries. However,
since we have no control of the phases present in split fermion models, we
can't
address the new contributions to CP-violating observables in a model
independent fashion. Nonetheless, there are
tree-level strong coupling contributions to these decays, so we
can expect significant contributions to the branching fractions in split
fermion models.

\begin{figure}[t]
\centerline{
\includegraphics[width=12cm,angle=0]{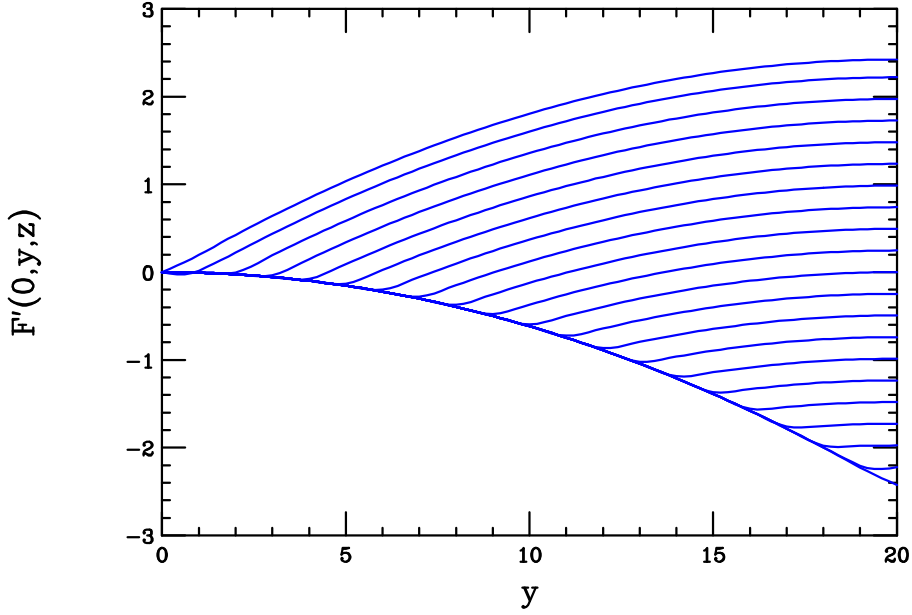}}
\caption{\small
KK sum for processes involving a single flavor changing vertex as a
function of one of the fermion positions, corresponding to the case where
one of the fermions is localized at the orbifold fixed point. The position
of the third (flavor conserved) fermion, $z$, is varied in steps of one
unit, with $z=0$ corresponding to the top curve and $z=20$ to the
bottom. The sum is done for a single extra dimension of size 20 units.
}
\label{fig:1fvc}
\end{figure}

The effective Lagrangian for a process with a single flavor change is
given by
\begin{gather}
{\cal L}_{\rm eff}^{\Delta b=1} = 
\frac{2}{3} g_s^2 \sum_{n=1}^{\infty} \frac{1}{M_n^2}
\sum_{i,j=L,R} U_{i(qb)}^{d(n)} \bar q_i \gamma^{\mu} b_i
\bar q_j \gamma^{\mu} q_j.
\end{gather}
Since one vertex is flavor diagonal, the sum depends on the absolute
position of the fermions. For instance, the analog of
Eq3. (\ref{eq:fnotsummed}-\ref{eq:f}) is
\begin{align}
F'(x,y,z) & = \sum_{n=1}^{\infty}
\frac{\cos(n\pi z)(\cos(n\pi x)-\cos(n\pi y))}{n^2}\notag\\
& = -\frac{\pi^2}{4} \left(\vphantom{\frac{1}{2}}|z+x| + |z-x| - |z+y| - |z-y| + \pi x^2 -
\pi y^2\right).
\end{align}
where $z$ corresponds to the location of the quark at the flavor
conserving vertex. This additional complication turns out to be minor, as
the actual
magnitude of the sum is similar to that in the previous case and does not
vary much over the parameter space, as can be
seen from Fig. \ref{fig:1fvc}.

We consider the decay amplitude
\begin{gather}
{\cal A}(B \to \phi K_S) = \frac{2}{3}g_s^2 R^2 
\sum_{i,j=L,R}F'(x_{b_i},x_{s_i},x_{s_j})
\langle \phi K_S|(\bar s_i \gamma^\mu b_i)(\bar s_j \gamma_\mu s_j)
|B^0\rangle 
\end{gather}
The relevant matrix elements are \cite{Barbieri:1997kq}
\begin{align}
\langle \phi K_S|(\bar s_L \gamma^\mu b_L)(\bar s_L \gamma_\mu s_L)
|B^0\rangle 
& = \frac{1}{3}H\notag\\
\langle \phi K_S|(\bar s_{L,i} \gamma^\mu b_{L,j})(\bar s_{L,j} \gamma_\mu
s_{L,i}) 
|B^0\rangle 
& = \frac{1}{3}H\\
\langle \phi K_S|(\bar s_L \gamma^\mu b_L)(\bar s_R \gamma_\mu s_R)
|B^0\rangle 
& = \frac{1}{4}H\notag\\
\langle \phi K_S|(\bar s_{L,i} \gamma^\mu b_{L,j})(\bar s_{R,j} \gamma_\mu
s_{R,i})
|B^0\rangle 
& = \frac{1}{12}H.\notag
\end{align}
Here, $i,j$ are color indices, displayed explicitly in the non-singlet
terms which now contribute. The common factor is
\begin{gather}
H = 2(\epsilon_\phi\cdot p_B)f_\phi m^2_\phi F_+(m_\phi^2),
\end{gather}
where $\epsilon_\phi$ represents the polarization vector of the phi meson,
and $F_+(q^2)$ is the form factor for this decay.
There are an additional four matrix elements obtained by taking
$(L\leftrightarrow R)$. We use the values $f_\phi = 233 \mev$
\cite{Gunion:1989we}, $F_+(m_\phi^2) = 0.38$ \cite{Wirbel:1985ji}, and $m_\phi
= 1020 \mev$ \cite{Hagiwara:2002fs}.

For the branching fraction we obtain (ignoring the {\cal O}(1) differences
among matrix elements)
\begin{align}
{\cal B}(B \to \phi K_S) & = \frac{1}{64 \pi m_B\Gamma_B}
f_\phi^2 m_\phi^4 
F_+^2(m_\phi^2) \left(\sum_{n=1}^{\infty} \frac{U^{d\dagger}_{L(ss)}
U^d_{L(bs)}}{n^2}\right)^2\\
& \approx
9.1\times 10^{-6}
\left(\frac{V_{sb,ss}^4}{0.04}\right)^2
\left(\frac{R}{1 \tev^{-1}}\right)^4
\left(
\sum_{i,j=L,R}F'(x_{b_i},x_{s_i},x_{s_j})
\right)^2,\notag
\end{align}
and similarly
\begin{gather}
{\cal B}(B \to \psi K_S)
\approx
5.6\times 10^{-5}
\left(\frac{V_{sb,cc}^4}{0.04}\right)^2
\left(\frac{R}{1 \tev^{-1}}\right)^4
\left(
\sum_{i,j=L,R}F'(x_{b_i},x_{s_i},x_{c_j})
\right)^2.
\end{gather}
Demanding that this not be larger than the observed rate gives the
approximate constraints $1/R \ge 1.0\ \tev$ from $B \to \phi K_S$ and $1/R
\ge 0.5\ \tev$ from $B \to \psi K_S$. These are not competitive with those
from $B_d$ and $B_s$ meson
oscillation, and provide a good consistency check. It is interesting to
note that if
a way can be found to reduce the constraints from the Kaon sector to the
few \tev\ scale
without disturbing the $b$-quark couplings (say by arranging the mixing
angles), then this contribution to $B \to \psi K_S,\phi K_S$ is of roughly
the same order as that of the Standard Model. Any new phases in this
scenario will thus contribute to the CP-violating observables with equal
effects in each decay channel.

We have also estimated the ree-level KK contribution to single top quark production
at LEP, $e^+ e^- \to \bar t q + t \bar q$ (with $q = u,c$) which proceeds
via KK $\gamma^{(n)}$ and $Z^{(n)}$ exchange. Using the parameterization
in \cite{Han:1998yr} we find an
effective anomalous flavor-changing vector coupling $v_Z$ given in this
case by
\begin{align}
v_{Z{\rm eff}} & = \left(
\sqrt{2} + 2 \sin\theta_W Q
\right) M_W^2 R^2
\sum_{i,j=L,R}\sum_{n=1}^{\infty} \frac{U^{u}_{i(tc)}U^{l}_{j(ee)}}{n^2}\\
& \approx
6.9 \times 10^{-4} 
\left(\frac{R}{1 \tev^{-1}}\right)^2
\sum_{i,j = L,R}
\left( \frac{V_{ut,ee}^4}{0.04}\right)
\left( 
\sum_{i,j=L,R}F'(x_{c_i},x_{t_i},x_{e_j})
\right),
\end{align}
where we have used only the $c \to t$ since it dominates the $u\to t$ by a
factor of 10. The current constraint on this effective flavor changing coupling from LEP
data is $v_Z \le 0.75$ 
\cite{Achard:2002vv}, and so split fermions in the up-quark sector are not
significantly constrained.

Lastly, we examine the potential gauge KK contributions to the purely
leptonic decay of neutral $B$ mesons, $B^0_q\to\ell^+\ell^-$, with
$q=d,s$.  This process proceeds through tree-level $\gamma^{(n)}$ and
$Z^{(n)}$ KK exchange, and could ,in principle, probes fermion splitting in the
leptonic sector as well as in the quark sector.  Here, for simplicity, we
will assume that the leptonic fields are all localized at the same
point in the extra dimension and will ignore any possible flavor
changing leptonic interactions.  The $\Delta b=1$ Lagragian describing
this decay is then
\begin{equation}
{\cal L}^{\Delta b=1}_{\rm eff}  =  \sum_{\alpha=\gamma,Z} 2G^2_\alpha
\, \sum_{\vec n=1}^\infty {\frac{1}{M_n^2}}\,\,\sum_{i=L,R} U^{d(n)}_{i(qb)}
g^b_{i,\alpha} \bar q_i\gamma^\mu b_i\, g^\ell_{i,\alpha}
\bar\ell_i\gamma_\mu\ell_i\,,
\end{equation}
where $G_\gamma=e$ and $G_Z=g/\cos\theta_w$.  
The hadronic matrix element governing this decay is given by
\begin{equation}
\langle 0|g^b_{L,\gamma/Z}U^{d(n)}_L\bar q_L\gamma^\mu b_L
         +g^b_{R,\gamma/Z}U^{d(n)}_R\bar q_R\gamma^\mu b_R| B^0_q\rangle
=if_bp^\mu_B[g^b_{L,\gamma/Z}U^{d(n)}_L-g^b_{R,\gamma/Z}U^{d(n)}_R]\,,
\end{equation}
where $p^\mu_B$ represents the momentum of the $B^0$ meson.  
Due to parity, only the axial-vector current contributes to this matrix 
element.  In the case of photon KK exchange, such contributions are
generated when the left- and right-handed fermions are localized at
separate points.  Here, we will ignore this possibility and consider
the case where only the $Z^{(n)}$ exchange mediates this decay.
The branching fraction is then
\begin{eqnarray}
{\cal B} (B_s\to\mu^+\mu^-)
 & = & {\frac{4G_F^2M_W^4f_B^2m_\ell^2m_B}{\pi c_w^4}}\tau_B R^4
\left[ 1-{\frac{4m_\ell^2}{m_B^2}} \right]^{1/2}\Big| \sum_n
\frac{g^b_{L,Z}U^{d(n)}_L-g^b_{R,Z}U^{d(n)}_R}{n^2} 
\Big|^2 \nonumber\\
& = & 1.15\times 10^{-6} \left( {\frac{R}{1\, {\rm TeV}^{-1}}}\right)
{\frac{V_{db,\mu\mu}^4}{(0.04)^2}} \nonumber\\
& & \quad \times \Big| g_{L,Z}F'(x_{b_L},x_{q_L},x_{\ell_L})
- g_{R,Z}F'(x_{b_R},x_{q_R},x_{\ell_R})\Big|^2 \,,
\end{eqnarray}
where we have chosen $q=s$ and used $f_B=200$ MeV.
The function $F'$ is as given in Eq. (19) and has a value in the
range $-2$ to $+2$ as shown in Fig. 8.  Taking values of this function which
maximizes the sum over the KK states, 
\ie, $F'=2$, yields a value of unity for the sum,
resulting in a branching fraction
of $1.15\times 10^{-6}$ for $R= 1\, {\rm TeV}^{-1}$.  This is a significant
enhancement over the Standard Model value\cite{Buchalla:1996vs} of
${\cal B}(B_s\to\mu^+\mu^-)\simeq 4.0\times 10^{-9}$.  The experimental
bound on this decay, ${\cal B}(B_s\to\mu^+\mu^-)< 2.6\times 10^{-6}$, as
determined by CDF\cite{Abe:1998ah}, sets the limit
$R^{-1}> 815$ GeV when the sum over the KK states takes on its maximal
value.  The sensitivity of this decay mode in probing the size of the
additional dimension is thus comparable to that of $B_s$ mixing and will
improve with Run II data at the Tevatron\cite{Anikeev:2001rk}.

\section{Conclusions}

Models where the Standard Model fermions are localized at
specific points along a compact extra dimension offer an
attractive means for constructing the fermion mass hierarchy
and suppressing dangerous operators such as proton decay.
In these scenarios, the fermions obtain narrow Gaussian
wavefunctions in the additional dimension with a width
much smaller than the compactification scale.  The fermion 
Yukawa couplings are then generated
by the overlap of the localized wavefunctions for the left- 
and right-handed fermions.  Lighter fermions are thus more
widely separated than heavier ones.

The gauge fields are free to propagate throughout the bulk in
these scenarios and their KK excitations develop tree-level
flavor changing interactions which are proportional to the
overlap of their wavefunctions with those of the localized
fermions.  Gluons, as well as the electroweak gauge bosons,
then mediate flavor changing neutral current processes at
dangerous levels.  Previously, it was thought that the only
way to avoid stringent bounds on the size of the compact
dimensions was to minimize the separation of the fermion
fields, thus endangering the scenario's natural explanation
of the fermion hierarchy.

In this paper, we have reinvestigated these new FCNC
interactions and have performed a general, systematic, model
independent analysis.  Our results hold for any such
model of the fermion hierarchy with specific fermion 
geographies.  We have employed a model parameterization
which contains only three parameters: the size of the
extra dimension $R$, the scaled width of the localized
fermion $\rho=\sigma/R$, and the fermion separation 
distance expressed in units of the width, $\Delta x =
\alpha\sigma$.  We performed a simple Monte Carlo analysis
and determined that the fermion mass hierarchy can be
reproduced in our parameterization for natural values of
the parameters.

We then evaluated the KK gluon tree-level flavor changing 
contributions to neutral meson oscillations.  We found
that the sum over the KK states is exponentially damped
for higher KK gauge states as the KK states can then
resolve the finite size of the fermion wavefunction.
This allows us to perform the KK sum in a scenario with more than one extra
dimension.  We then evaluated the constraints from Kaon
mixing in the case of one extra dimension and confirmed 
previous results that $1/R\gsim 100$'s 
TeV for larger values of $\rho$ unless the separation was 
very small.  However, the constraint shrinks to $1/R\gsim$
few TeV for smaller values of $\rho$, even for widely
separated fermions, at the expense of introducing a new hierarchy between
the compactification and fermion localization scales.  The constraints
from $B_d$ and $B_s$
mixing were found to be much less restrictive.  We also
performed the evaluation for the case of two or more 
additional dimensions and found that the FCNC constraints
were much more difficult to evade.  

We next studied the dependence of our constraints on the
fermion mass mixing matrices, and found that with a
realignment of the matrix elements our bounds could be
reduced further by factors of 10-100.

In addition, we examined
the rare meson decays $B_d\to\psi K_S,\phi K_S$, as well
as single top-quark production in $e^+e^-$ collisions, 
and found weaker limits of the size of the extra dimension
of order TeV$^{-1}$.  We note that the KK gluon contributions
to these rare decays are significant enough to generate
interesting effects in the related CP violation observables.

In summary, we have shown that once the parameter space
is systematically explored, it is possible to evade the
stringent bounds from FCNC in split fermion models for
natural values of the parameters and without the introduction
of any additional hierarchies.  Lastly, we note that the
introduction of brane localized kinetic terms are known
to significantly reduce the couplings of gauge KK states
\cite{Carena:2002me,delAguila:2003bh} and may help to even further reduce the
constraints from FCNC in these scenarios.

\section*{Acknowledgements}
%\noindent{\bf Acknowledgements}

The authors would like to thank Tom Rizzo, Frank Petriello, and Jeremy
Copeland for helpful discussions.

\section{Appendix}

Here we present a cute way to perform the sum over KK modes analytically in
one dimension, and see that the sum is exactly linearly proportional to the
separation.\cite{jeremy} The functions that we need are
\begin{gather}
F(x,y) \equiv \sum_{n=1}^\infty \frac{(\cos(n\pi x)-\cos(n\pi y))^2}{n^2},
\end{gather}
and
\begin{gather}
G(x_1,x_2,y_1,y_2) \equiv \sum_{n=1}^\infty \frac{(\cos(n\pi
x_1)-\cos(n\pi y_1))(\cos(n\pi x_2)-\cos(n\pi y_2))}{n^2}.
\end{gather}
We can do both of these by evaluating
\begin{gather}
f(x) \equiv \sum_{n=1}^\infty \frac{\cos(nx)}{n^2}.
\end{gather}
So that
\begin{align}
F(x,y) & = \frac{1}{2}(f(2\pi x) + f(2\pi y) + \frac{\pi^2}{3} - 2f(\pi x
+ \pi y) - 
2f(\pi x - \pi y))\label{eq:bigf}\\ 
G(x_1,x_2,y_1,y_2)
& = \left[\vphantom{\frac{1}{2}}f(\pi x_1 + \pi x_2) + f(\pi x_1 - \pi
x_2) +f(\pi y_1 + \pi y_2) + f(\pi y_1 - \pi y_2)\right.\notag\\
&\left. \vphantom{\frac{1}{2}}- f(\pi x_1 + \pi y_2) - f(\pi x_1 - \pi
y_2) - f(\pi y_1 + \pi x_2) - f(\pi y_1 - \pi x_2)\right]\label{eq:bigg}
\end{align}
Writing the $\cos(nx)$ as two exponentials and combining the sums we have
\begin{gather}
f(x) = \frac{1}{2}\sum_{n\ne0} \frac{e^{inx}}{n^2}.
\end{gather}
The function $f$ is then the solution of the differential equation
\begin{align}
f''(x) & = -\frac{1}{2}\sum_{n\ne 0}e^{inx}
= -\frac{1}{2}\sum_{n=-\infty}^{\infty}e^{inx} + \frac{1}{2}\\
& = -\frac{1}{2}\left(2\pi \sum_{k=-\infty}^{\infty}
\delta(x-2\pi k) - 1\right).
\end{align}
This is solved by
\begin{gather}
H(x) =
-\frac{1}{2}\left(\pi|x|-\frac{x^2}{2}-\frac{\pi^2}{3}\right),
\end{gather}
where $H$ is defined on the interval $[-\pi,\pi]$ and is $2\pi$-periodic
for other values (this takes care of the sum over delta functions). We
then have
\begin{gather}
f(x) = H(x) + \alpha x + \beta.\label{eq:littlef}
\end{gather}
Looking at the original function we see that we must have
$\int_{-\pi}^{\pi}f = 0$ which gives $\beta = 0$. Also, since $f$ is an
even function $\alpha = 0$.

We can then use Eq. (\ref{eq:littlef}) in (\ref{eq:bigf}) and
(\ref{eq:bigg}) and use the physical condition that all arguments are
positive to get the final result
\begin{gather}
F(x,y) = 
\frac{\pi^2}{2}|x-y|,
\end{gather}
and
\begin{gather}
G(x_1,x_2,y_1,y_2) = 
-\frac{\pi^2}{4}
\left(\vphantom{\frac{1}{2}}|x_1-x_2| + |y_1 - y_2| - |x_1 - y_2| - |y_1 -
y_2|\right). 
\end{gather}

\bibliography{fcnc}

\end{document}